\documentclass[12pt,preprint]{aastex}

\shorttitle{} \shortauthors{An, Goss, Zhao et al.}
\begin{document}

\title{Simultaneous Multi-Wavelength Observations of Sgr~A*}

\author{T. An\altaffilmark{1},
W. M. Goss\altaffilmark{2}, Jun-Hui Zhao\altaffilmark{3}, X. Y.
Hong\altaffilmark{1}, S. Roy\altaffilmark{4}, A. P. Rao \altaffilmark{4} and
Z.-Q. Shen\altaffilmark{1} }

\altaffiltext{1}{Shanghai Astronomical Observatory, Chinese Academy of
Sciences, Shanghai 200030, PR China; antao@shao.ac.cn}

\altaffiltext{2}{National Radio Astronomy Observatory, P. O. Box O, Socorro, NM
87801; }

\altaffiltext{3}{Harvard-Smithsonian CfA, 60 Garden St, MS 78, Cambridge, MA
02138; }

\altaffiltext{4}{National Center for Radio Astrophysics (TIFR), Pune University
Campus, Post Bag No. 3, Ganeshkhind, Pune 411 007, India;}

\begin{abstract}
We observed Sgr A* using the Very Large Array (VLA) and the Giant Metrewave
Radio Telescope (GMRT) at multiple cm and mm wavelengths on 17 June 2003. The
measured flux densities of Sgr A*, together with those obtained from the
Submillimeter Array (SMA) and the Keck II 10 m telescope on the same date, are
used to construct a simultaneous spectrum of Sgr A* from 90 cm to 3.8 $\mu$m.
The simultaneous spectrum shows a spectral break at about 3.6 cm, a possible
signature of synchrotron self-absorption of the strong radio outburst which
occurred near epoch 17 July 2003. At 90 cm, the flux density of Sgr~A* is
$0.22\pm0.06$ Jy, suggesting a sharp decrease in flux density at wavelengths
longer than 47 cm. The spectrum at long cm wavelengths appears to be consistent
with free-free absorption by a screen of ionized gas with a cutoff $\sim$100
cm. This cutoff wavelength appears to be three times longer than that of
$\sim$30 cm suggested by Davies, Walsh \& Booth based on observations in 1974
and 1975. Our analysis suggests that the flux densities of Sgr~A* at
wavelengths longer than 30 cm could be attenuated and modulated by stellar
winds from massive stars close to Sgr~A*.
\end{abstract}


\section{INTRODUCTION} \label{intro}

Soon after the discovery of the compact radio source Sagittarius A* (Sgr~A*) at
the Galactic center (GC) \citep{BB74}, Davies, Walsh \& Booth (1976, hereafter
DWB) carried out observations of Sgr~A* at 0.408, 0.96 and 1.66 GHz. Sgr~A* was
only detected at the two higher frequencies. The measurements showed a
low-frequency turnover around 1 GHz in the spectrum of Sgr~A*.
The authors attributed the decrease in flux density below 1 GHz to free-free
absorption from ionized gas of Sgr A West. Recent observations of Sgr~A* at
both 620 and 330 MHz indicate that the spectrum below 1 GHz may indeed have a
pronounced turnover. At 620 MHz, Roy \& Rao (2004) detected Sgr~A* using the
GMRT with a flux density of $0.5\pm0.1$ Jy with an angular resolution of
$11.4\arcsec\times7.6\arcsec$, while \cite{Nord04} detected Sgr~A* at 330 MHz
using the VLA of the National Radio Astronomy Observatory (NRAO) with a flux
density of $0.33\pm0.12$ Jy (resolution of $\sim$9\arcsec).

Sgr~A* is now believed to be associated with a supermassive black hole (SMBH)
with a mass $\sim4\times10^6 M_{\odot}$ \citep*{Schodel02,Ghez03}. A number of
models have been proposed to explain the observed phenomena in the vicinity of
Sgr~A* at radio, near-IR and X-ray wavelengths (Melia \& Falcke 2001). Coker \&
Melia (1997) proposed that stellar winds, which originate from massive stars
near Sgr~A*, might play an important role in fuelling Sgr~A* and therefore may
have caused the variations in flux density at radio wavelengths. These models
are sensitive to the observed spectrum of Sgr~A*. Simultaneous multi-wavelength
observations could provide important constraints for current accretion or
outflow theories. If stellar winds are indeed responsible for the flux density
variations at millimeter/submillimeter wavelengths, a cutoff at long radio
wavelengths could be sensitive to changes in stellar wind flux near Sgr~A*.

To determine the low-frequency turnover in the spectrum of Sgr~A*, we have
performed multi-wavelength observations using the VLA at 90, 26, 23, 20, 17, 6,
3.6, 2, 1.3 and 0.7 cm, and the GMRT at 47 cm on 17 June 2003 (UT). At the same
epoch, both the SMA (Moran 1998; Ho, Moran \& Lo 2004) and the Keck II 10 m
telescope observed Sgr~A* at 0.89 mm and 3.8~$\mu$m, respectively. In this
letter, we present results from the quasi-simultaneous observations.

\section{OBSERVATIONS AND DATA REDUCTION}\label{obser}

The VLA and GMRT observations are summarized in Table \ref{tab:obs}, including
observing band ($\lambda$), frequency ($\nu$), bandwidth ($\Delta\nu$),
on-source time ($t$) and flux density of the primary calibrator ($S_{c}$). The
observations with the VLA were carried out in the A configuration at
wavelengths from 90 cm to 7 mm. The observations at 90 and 20 cm with the VLA
and at 47 cm with the GMRT were performed in spectral-line modes in order to
reject radio frequency interference and minimize bandwidth smearing. The raw
data were reduced using the Astronomical Image Processing System (AIPS) of the
NRAO following standard procedures. Absolute flux density calibration was
performed using 3C~286 (at 0.7, 1.3 and 2 cm) or 3C 48 (at other wavelengths)
following the flux density scales of Baars et al. (1977). The flux density
calibration of 3C~48 or 3C~286 was carried out using the visibilities over
appropriate {\it uv} ranges suggested by VLA documentation at wavelengths
between 90 and 2 cm; clean models of 3C 286 (obtained from C.Chandler) were
used for flux density calibrations at 1.3 and 0.7 cm using all visibilites. A
few nearby QSOs were interleaved during the observations of Sgr~A* to determine
the complex gains. An amplitude and phase calibration of 3C~48 or 3C~286 was
performed using a two-minute solution interval to obtain average flux density
scales over the scan, which were then applied to the phase calibrators and Sgr
A*. A few iterations of phase self-calibrations of calibrated \emph{uv} data
were also performed to correct for residual phase errors. The GMRT calibration
procedure for observation of the Galactic center is described in detail by Roy
\& Rao (2004). The primary flux density calibrator was 3C~48, whose flux
density is 28.5 Jy at 640 MHz (Baars scale). Corrections were made for the
increase of the system temperature in the direction of the Galactic center
compared to the high galactic latitude field of 3C~48. The point sources
1751$-$253 and 1741$-$25 were used as secondary phase and bandpass calibrators,
respectively.

\section{DATA ANALYSIS AND RESULTS}

The flux densities of Sgr~A* were determined in the image domain. Either a
point-source or an elliptical Gaussian model was used to fit the data. The flux
density measurement at 90 cm will be discussed in Section \ref{Detection}. At
20 cm, Sgr~A* is slightly resolved; thus a 2-D elliptical Gaussian function was
used to fit the long baseline ($>$20 k$\lambda$) visibilities. At wavelengths
between 6 and 0.7 cm, free-free emission from the HII regions in Sgr A West is
resolved out on baselines longer than 80 k$\lambda$.
The flux density of a point source can be directly determined in the \emph{uv}
domain by averaging the amplitude of the visibilities on baselines
$>80k\lambda$. The difference in flux densities between the two procedures is
consistent with the uncertainty (${\rm \sigma_M}$) in the measurements. The
calibration uncertainty ($\rm \sigma_C$) is dominated by the uncertainty in the
calibration of the flux density scale. The final 1$\sigma$ errors are the
quadrature addition of the two terms, \emph{i.e.} $\sqrt{\sigma_M^2 +
\sigma_C^2}$. The measurements of flux densities at 2, 1.3 and 0.7 cm are
consistent with the values derived from the VLA monitoring program
\citep{HZBG04}. The GMRT image at 47 cm was made with a \emph{uv} range 10--50
k$\lambda$ to minimize confusion from the Sgr A complex. A Gaussian model was
fitted to the flux density distribution of Sgr A* and a constant term with a
slope was fitted to the background using JMFIT in AIPS. The flux density of Sgr
A* is 0.45$\pm$0.10 Jy at 47 cm.

\subsection{Flux density measurement of Sgr~A* at 90 cm}\label{Detection}

\begin{figure}\centering
\includegraphics[width=0.4\textwidth]{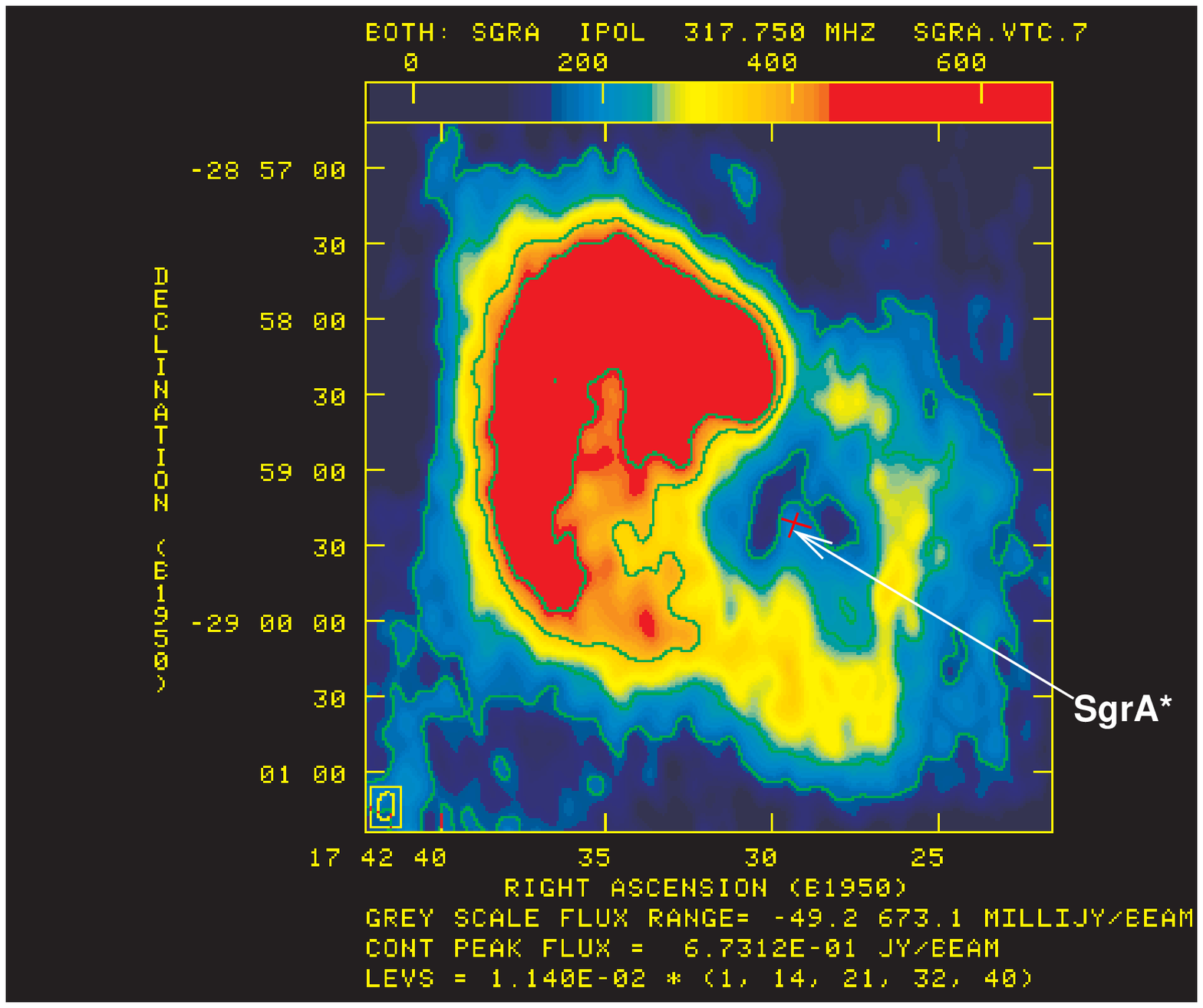}
\caption{Pseudo-color/contour image of Sgr~A* at 90 cm.
Contours: $11.4\times(1,14, 21, 32, 40)$ mJy/beam. In order to compare our
results with those obtained by \cite{Nord04} at the same wavelength, we have
restored the final image using a similar synthesized beam, \emph{i.e.},
$10.9''\times6.8''$ (PA$=-10\degr$). The \emph{r.m.s.} noise in the image is 12
mJy/beam
. The red cross at the
image center marks the expected scattering size and position of Sgr~A* at 90
cm, \emph{i.e.}, $\sim$$11''\times6''$ at PA= 80\degr{}
\citep{Rogers94,LSZH98}. }\label{fig:sgra90}
\end{figure}

\begin{figure}\centering
\includegraphics[width=0.5\textwidth]{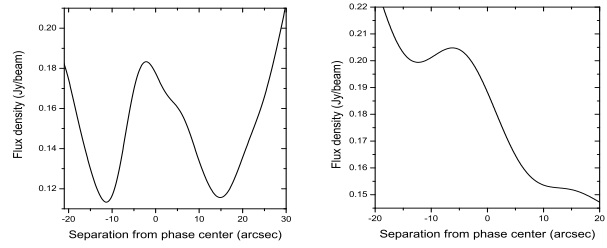}
\caption{Cross cuts along the expected axes of Sgr~A* at 90~cm. The slices are
centered at 17$^h$42$^m$29.4$^s$, $-28\degr59'18''$ (B1950). \emph{Left:} Slice
through the major axis (PA$=80\degr$) with a resolution of 6.8\arcsec. East is
to the left; \emph{Right:} Slice through the minor axis (PA$=-10\degr$) with a
resolution of 10.9\arcsec. South is to the left. }\label{fig:slice}
\end{figure}

Figure \ref{fig:sgra90} shows the central $4.5\arcmin$ region of the Galaxy at
90 cm. The supernova Sgr A East dominates the total flux density at 90 cm. An
excess in flux density is observed at the expected location of Sgr~A* (red
cross at the image center in Figure \ref{fig:sgra90}); the region surrounding
Sgr~A* is significantly absorbed by the ionized gas in Sgr A West, except for
the extended emission south of Sgr~A*. We made two slices along the major
(PA$=80\degr$) and minor (PA$=-10\degr$) axes of the expected scattering shape
to determine the flux density of Sgr~A* (Figure~\ref{fig:slice}). Gaussian
profiles were fitted to each of the slices after subtracting diffuse emission
from the background; an apparent size of
$(16.0\arcsec\pm1.3\arcsec)\times(15.3\arcsec\pm2.5\arcsec)$ was determined.
The errors in the parameters are determined using the 2-D Gaussian fitting
procedure JMFIT in AIPS and also checked by fitting the two 1-D slices
(Figure~\ref{fig:slice}). The source size of Sgr~A* at 90 cm is
$(14.4\arcsec\pm1.4\arcsec)\times(10.7\arcsec\pm2.4\arcsec)$
and ${\rm PA=95\pm25\degr}$, after deconvolution. The fitted peak flux density
is $65\pm16$ mJy/beam, with a total flux density of $220\pm60$ mJy. The
observed flux density and source size of Sgr~A* at 90 cm agree with the
determinations of \cite{Nord04} within 1$\sigma$.

\subsection{Simultaneous Spectrum on 17 June 2003}

\begin{figure}\centering
\includegraphics[width=0.45\textwidth]{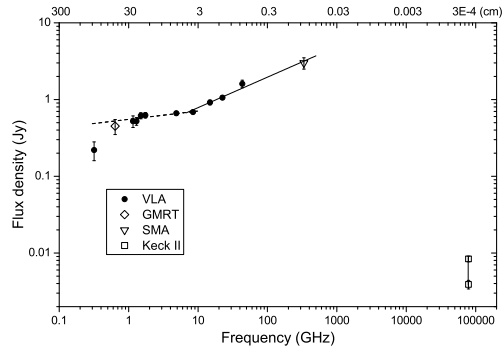}
\caption{Simultaneous spectrum of Sgr~A* from 90 cm to 3.8 $\mu$m on 17 June
2003. \emph{solid circle}: VLA; \emph{diamond}: GMRT; \emph{triangle}: SMA
(private communication, J.-H. Zhao); \emph{square}: Keck II \citep{Ghez04}. }
\label{fig:spec}
\end{figure}

Figure~\ref{fig:spec} shows the quasi-simultaneous spectrum of Sgr~A* at
wavelengths ranging from 90 cm to 3.8 $\mu$m, observed on 17 June 2003.
Table~\ref{tab:spec} summarizes the flux density measurements. A rising
spectrum extends from  $\sim$2 cm to the submillimeter band. However, the
wavelength of the peak emission is quite uncertain since there are no
measurements from 0.89 mm to 3.8 $\mu$m. The rising spectrum from the short cm
to mm wavelengths can be described by a power law ($S_\nu\propto\nu^{\alpha}$)
with $\alpha^{3.6cm}_{0.89mm}=0.43\pm0.04$ ({\it solid line} in Figure
\ref{fig:spec}); the spectrum between 3.6 and 47 cm is flatter,
$\alpha^{47cm}_{3.6cm}=0.11\pm0.03$ ({\it dashed line} in Figure
\ref{fig:spec}). The difference in spectral indices between the two parts of
the power-law spectra is
$\alpha^{3.6cm}_{0.89mm}-\alpha^{47cm}_{3.6cm}=0.32\pm0.05$, suggesting a
significant break near 3.6 cm on 17 June 2003.

A break in the cm-band spectrum of Sgr~A* has been suggested previously
\citep{ZML92,Falcke98,Zhao03}. The determined break wavelength of 3--4 cm is
consistent with the previously suggested break of 2--3.5 cm suggested by Falcke
et al. (1998). The excess in flux density towards short wavelengths in the
spectrum of Sgr~A* might be associated with activities in the inner region of
the accretion disk or jet nozzle (Melia \& Falcke 2001). In comparison with the
radio light curves, the increased spectral index at wavelengths shorter than
the break of $\sim$3.6 cm on 17 June 2003 is consistent with the fact that
Sgr~A* was undergoing a large radio outburst. The VLA monitoring campaign
(Herrnstein et al. 2004) did show that Sgr~A* was in an outburst stage at 7 mm
near the epoch of 16 June 2003; the outburst apparently peaked near the epoch
of 17 July 2003. The flux density at 7 mm on 17 July 2003 was at the highest
level between 21 June 2000 and 16 October 2003 (Herrnstein et al. 2004).
Therefore, a break in the spectrum of Sgr~A* at a short cm wavelength could be
a signature of synchrotron self-absorption \citep{KP69} of a relatively strong
radio outburst, as proposed to explain the 3 October 2002 X-ray event and
corresponding radio outburst \citep{Zhao04}.

At 90 cm, the measured flux density of 0.22$\pm$0.06 Jy is far below the
inferred 0.49 Jy extrapolated from the power law fit
$S_\nu\propto\nu^{0.11\pm0.03}$. This significant decrease ($>4\sigma$) in flux
density suggests that the spectrum of Sgr~A* on 17 June 2003 could have a
cutoff at wavelengths longer than 47 cm.

\section{Discussion}

The earlier Jodrell Bank measurements of Sgr~A* made by DWB showed a
low-frequency turnover at about 1 GHz. The flux densities of 0.26$\pm$0.03 Jy
at 0.96 GHz and 0.56$\pm$0.06 Jy at 1.66 GHz had been corrected for resolutions
of the interferometer and the known angular diameters of Sgr~A*. We note that
the angular diameters measured with the Jodrell Bank at 0.96 and 1.66 GHz are
consistent with the values extrapolated from the relationship between the
angular size and observing wavelength ({\it e.g.}, Lo et al. 1998; Bower et al.
2004).
For angular sizes of $1.5\arcsec$ at 0.96 GHz and $0.5\arcsec$ at 1.66 GHz,
Sgr~A* is dominant on baselines from 34 to 75 k$\lambda$ and from 60 to 132
k$\lambda$, respectively.
 The errors in flux densities at these two frequencies
($\sim$10\%) are consistent with those estimated from our current data at 20
cm. However, the 408 MHz upper limit measured by DWB with a single baseline is
quite uncertain because of confusion arising from the complex region in Sgr A
East in close proximity to the broadened source Sgr~A* at 408 MHz ({\it e.g.}
$\sim$8\arcsec, Roy \& Rao 2004). This upper limit
will be excluded in following discussion.

\begin{figure}\centering
\includegraphics[width=0.4\textwidth]{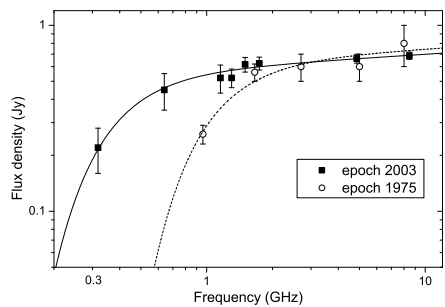}
\caption{Comparison of the low-frequency spectra of Sgr~A* observed between
epochs 2003 (\emph{solid square}) and 1975 (\emph{open circle}). The {\it
dashed} and {\it solid} lines represent fits using free-free absorption opacity
to the spectrum at epochs 1975 and 2003, respectively. The free-free opacity
decreases by a factor of $\sim$nine at 30 cm from 1975 to 2003.
}\label{fig:LFspec}
\end{figure}

Figure \ref{fig:LFspec} shows the observed spectra for Sgr A* at epochs 1975
(DWB) and 2003 (this Letter) for wavelengths longer than 3.6 cm. A ratio of
$2.0\pm0.4$ ($S_{1.16GHz}$(2003)/$S_{0.96GHz}$(1975)) would suggest a
significant increase in the flux density at about 1 GHz from epoch 1975 to
2003. The spectra at both epochs can be fitted with a model of a slowly rising
power law spectrum ($\propto\nu^{\alpha}$) together with a free-free absorption
screen ($e^{-\tau_{ff}}$) between Sgr~A* and the observer, as first proposed by
DWB.  If the flux density at 0.96 GHz measured by DWB is indeed reliable, then
the comparison of the fits to the 1975 (\emph{dashed line} in Figure
\ref{fig:LFspec}) and 2003 data (\emph{solid line} in Figure \ref{fig:LFspec})
does suggest a substantial change in the spectrum of Sgr~A* at wavelengths
longer than 30 cm in the past 28 years.

If the free-free absorption model is in fact correct, the variation in the
low-frequency spectrum would suggest that the column density of the free-free
absorption screen must have decreased significantly over the past 28 years.
Starting from DWB's free-free absorption model ($S_\nu\propto \nu^\alpha
e^{-\tau_{ff}}$), the quantitative change in the column density of ionized gas
in front of Sgr~A* can be assessed. The data observed at both epochs can be
fitted with a power-law spectrum $S_\nu\propto\nu^{0.074\pm0.022}$ at the
higher frequencies along with an exponential cutoff at the lower frequencies
arising from free-free absorption. The comparison of the model fits to the
observations at these two epochs suggests that free-free absorption opacity
($\tau_{ff})$ towards Sgr~A* must have decreased by a factor of nine from 1975
to 2003. The critical cutoff in wavelength due to free-free absorption has
changed from $\sim$30 cm to $\sim$100 cm. Assuming the electron temperature of
the ionized gas remains constant, the inferred decrease in $\tau_{ff}$ would
then correspond to a decrease by a factor of nine in emission measure $EM=\int
n_e^2 dl$. Such a large variation in EM is unlikely to arise from changes in
electron density in the halo of Sgr A East (Pedlar et al. 1989).

If the changes in EM and electron density do occur in a compact region within
0.1\arcsec--10\arcsec{} of SgrA* (0.004--0.4 pc), stellar winds which originate
from massive stars towards Sgr A* could provide the changing environment. For
example, Loeb (2004) suggested that stellar winds from the close-in stars could
modulate low-frequency flux density of Sgr~A* over a timescale of about ten
years through orbital motions of the stars. Here we consider two types of
stellar winds: (1) the stellar wind is isotropic ($n_e\propto r^{-2}$) and (2)
the stellar wind forms a one-dimensional flow due to the gravity of the SMBH
($n_e\propto r^{-1}$). Using the star S2 \citep{Schodel02} as an example, the
orbit period is about 15 years and the ratio of the pericenter and apocenter is
about 1:15; the star has a surface temperature of about 30000K, providing an
ionization zone of H roughly equivalent of a B0 star (Ghez et al. 2003). The
stellar wind mass flux at pericenter in case (1) is about 200 times that at
apocenter; in case (2) the ratio in stellar wind flux between pericenter and
apocenter is 15:1. The ratio of three in electron column density between 1975
and 2003 inferred from the change of the low-frequency cutoff is much less than
that of the stellar wind flux in both cases. Nevertheless, large amplitude
fluctuations of accreted matter within 0.002 pc (0.05\arcsec) captured from the
stellar winds near Sgr A* over a time scale of less than a few decades were
predicted in numerical simulations (Melia \& Coker 1999). A variation in the
radio continuum spectrum was also suggested by the simulations, although the
authors attributed the low-frequency turnover to refraction effects. It is more
likely that the source for the low-frequency turnover is free-free absorption
from cool gas in close proximity to Sgr~A*. The colder gas ($\sim10^4$ K) can
co-exist with the hot gas ($\sim10^7$ K) that provides the X-ray emission near
Sgr~A* (Cuadra et al. 2005a, 2005b). For a stellar wind with a size of 0.004
pc, the free-free absorption opacity is about unity at about 0.4 GHz for an
electron density of $10^4$ cm$^{-3}$ and electron temperature of $10^4$ K. The
variation of the low-frequency turnover due to free-free absorption towards
Sgr~A* appears to be sensitive to fluctuations of the stellar wind mass
captured by the SMBH potential.

\acknowledgments

We thank the referee and H. Falcke for helpful comments. The work is supported
by the NSFC (10503008, 10328306, 10333020). The VLA is operated by the National
Radio Astronomy Observatory, a facility of the National Science Foundation
operated under cooperative agreement by Associated Universities, Inc. The GMRT
is a component of the National Centre for Radio Astrophysics of the Tata
Institute of Fundamental Research.

\clearpage

\clearpage

\clearpage

\clearpage

\clearpage

\begin{table}\centering
\caption{OBSERVATIONAL DATA OF SGR~A* ON 17 JUNE 2003}
\begin{tabular}{ccccc}\hline\hline
$\lambda$&$\nu$ & $\Delta\nu$ & $t$ & $S_{c}$  \\
(cm) & (GHz) & (MHz) & (min) & (Jy) \\\hline
90 &0.303/0.332         & 6    &29 & 3C48(45.2/42.9)\\
47 & 0.640              & 16   &85 & 3C48(28.5) \\
20$^a$ &1.16/1.30/1.50/1.74 & 34 &56& 3C48(18.6/17.0/15.2/13.5)\\
6 & 4.835/4.885         & 100  &11& 3C48(5.5/5.4)\\
3.6 &8.435/8.485        & 100  &11& 3C48(3.2/3.1)\\
2 &14.915/14.965        & 100  &11& 3C286(3.5/3.5)\\
1.3 &22.435/22.485      & 100  &29& 3C286(2.52)  \\
0.7 & 43.315/43.365     & 100  &32& 3C286(1.45)  \\\hline
\end{tabular}\label{tab:obs}\\
\raggedright $^a$ observations at 20 cm were carried out at four frequencies.
\end{table}

\clearpage

\begin{table}
  \centering
  \caption{FLUX DENSITY DETERMINATIONS OF SGR A* ON 17 JUNE 2003}\label{tab:spec}
\begin{tabular}{ccccc}
  \hline\hline
$\lambda(cm)$ & S(Jy) & $\sigma_{T}(Jy)$ & $\sigma_C(\%)$& $\sigma_{M}(\%)$
\\\hline
  90 & 0.22 & 0.06 & 10 & 25 \\
  47 & 0.45 & 0.10 & 10 & 20 \\
  26 & 0.52 & 0.09 & 3  & 17 \\
  23 & 0.52 & 0.06 & 4  & 11 \\
  20 & 0.62 & 0.06 & 5  & 7 \\
  17 & 0.63 & 0.05 & 5  & 6 \\
   6 & 0.66 & 0.04 & 4  & 4 \\
 3.6 & 0.69 & 0.03 & 4  & 2 \\
 2   & 0.92 & 0.06 & 7  & 1 \\
 1.3 & 1.06 & 0.06 & 5  & 2 \\
 0.7 & 1.6  & 0.2  & 11 & 1 \\\hline
\end{tabular}
\end{table}

\end{document}